\documentclass[useAMS,usenatbib,onecolumn]{mn2e}
\usepackage{epsfig}
\usepackage{graphicx}

\newcommand{\etal}{{et~al.}~}
\newcommand{\lsim}{\,\lower2truept\hbox{${<\atop\hbox{\raise4truept\hbox{$\sim$}}}$}\,}
\newcommand{\gsim}{\,\lower2truept\hbox{${>\atop\hbox{\raise4truept\hbox{$\sim$}}}$}\,}
\newcommand{\beq}{\begin{equation}}
\newcommand{\eeq}{\end{equation}}
\newcommand{\COBE}{$COBE$-DMR}
\newcommand{\WMAP}{$WMAP$}
\newcommand{\fstica}{{\sc{fastica}}}

\newcommand{\vect}[1]{{\mathbfit{#1}}}

\def\aa{{\sl Astron.\ \&\ Astrophys.\ }}
\def\aas{{\sl Astron. \& Astrophys.\ Suppl.\ }}
\def\apj{{\sl Astrophys.\ J.\ }}

\def\apjs{{\sl Astrophys.\ J.\ Supp.\ }}

\def\ieeespl{{\sl IEEE\ Signal\ Processing\ Lett.\ }}

\def\mnras{{\sl MNRAS\ }}
\def\n{{\sl Nature\ }}

\def\nn{{\sl Neural Networks\ }}
\def\nar{{\sl New\ Astron.\ Rev.\ }}

\def\prd{{\sl Phys.\ Rev.\ D\ }}

\def\pieee{{\sl Proc.\ IEEE\ }}

\title[CMB signal in \WMAP~3yr data with \fstica]
{CMB signal in \WMAP~3yr data with \fstica}

\author[Maino \etal]{D. Maino$^1\footnote{E-mail:davide.maino@mi.infn.it}$,
S. Donzelli$^1$, A.~J. Banday$^2$, F. Stivoli$^{3}$, C. Baccigalupi$^{3,4}$\\
$^{1}$ Dipartimento di Fisica, Universit\`a di Milano, Via Celoria 16, I-20133, Milano, Italy\\
$^{2}$ Max-Planck Institute f{\"u}r Astrophysik, Karl-Schwarzschild Str. 1, D-85740, Garching, Germany\\
$^{3}$ SISSA/ISAS, Astrophysics Sector, Via Beirut 4, I-34014, Trieste\\
$^{4}$ INFN, sezione di Trieste, Via Valerio 2, I-34127, Trieste, Italy
}

\pagerange{\pageref{firstpage}--\pageref{lastpage}}
\pubyear{2006}

\begin{document}

\maketitle

\label{firstpage}

\begin{abstract}
We present an application of the fast Independent Component Analysis (\fstica)
to the \WMAP~3yr data with the goal of extracting the CMB signal. We 
evaluate the confidence of our results by means of Monte Carlo simulations 
including CMB, foreground contaminations and instrumental noise specific 
of each \WMAP~frequency band.
We perform a complete analysis involving all or a subset of the 
\WMAP~channels in order to select the optimal combination for CMB 
extraction, using the frequency scaling of the reconstructed component as 
a figure of merit. We found that the combination KQVW provides the best 
CMB frequency scaling, indicating that the low frequency foreground 
contamination in Q, V and W bands is better traced by the emission in 
the K band. \\
The CMB angular power spectrum is recovered up to the degree scale, 
it is consistent within errors for all \WMAP~channel combination considered, 
and in close agreement with the \WMAP~3yr results. \\
We perform a statistical analysis of the recovered CMB pattern, and 
confirm the sky asymmetry reported in several previous works with 
independent techniques.
\end{abstract}

\footnotetext{E-mail: davide.maino@mi.infn.it}

\section{Introduction}
\label{intro}

The Cosmic Microwave Background (CMB) anisotropies are at present the main 
tracer of the physical processes occurred in the very early universe; a complete 
mapping of them may reveal crucial clues about the statistics of the primordial 
perturbations, the existence of gravity waves, the global geometry of the 
universe as well as the physical components responsible for the accelerated 
expansion eras in the very early and recent universe, known as inflation and dark 
energy, respectively. \\
For these reasons, a great experimental work is being carried out in order to 
measure the finest structure of the CMB anisotropies. In total intensity, 
right after successful sub-orbital observations, the Wilkinson Microwave Anisotropy 
Probe (\WMAP) satellite is performing all sky observations of the CMB in five frequency 
bands ranging between 22 and 90~GHz (see \citealt{spergel_etal_2006} and references therein). 
The first detections of the CMB polarization have been made from the ground 
\citep{kovac_etal_2002}, balloons \citep{montroy_etal_2005} as well as from 
the \WMAP~satellite itself \citep{page_etal_2006}. In a few years the {\sc Planck} 
satellite\footnote{http://www.rssd.esa.int/Planck} will be launched, performing all 
sky measurements of the CMB total intensity and polarization anisotropy 
in 9 frequency bands between 30 and 857~GHz, with an angular resolution 
reaching 5 arcminutes with a sensitivity of a few micro-Kelvin. Next generation 
sub-orbital and satellite observations are ongoing or planned\footnote{http://lambda.gsfc.nasa.gov/}. 

The observations are confirming that the Galactic and extra-Galactic foreground 
emissions represent one of the two most challenging obstacles to the complete 
knowledge of CMB anisotropies. 
The other one is represented by the control of instrumental systematics. 
In total intensity the sky at high Galactic latitudes appears dominated 
by the CMB emission at least in the frequency range between 60 and 90~GHz 
observed by \WMAP, although residual foreground contamination might be 
responsible for the deviations of the observed signal from a Gaussian 
statistics (see \citet{copi_etal_2006} and references therein). 
In polarization the sky remains foreground dominated even 
if the region containing the brightest Galactic emission is cut out 
\citep{page_etal_2006}. In particular, foregrounds are likely to be 
comparable or higher to the the curl component in the CMB 
polarization at all frequencies, and in any region of the sky 
\citep{baccigalupi_2003}. \\
For these reasons it is important to study and test data analysis 
algorithms which are able to remove foreground contamination 
from the data. The class of these tools is called component 
separation, and they all make use of the multi-frequency superposition 
of the background and foreground emissions in order to 
separate them out. The concepts and means they use, however, are 
very different. The category of non-blind methods uses prior 
knowledge of the foreground emission in order to recover the CMB 
component \citep{brandt_etal_1994,tegmark_efstathiou_1996,hobson_etal_1998,bouchet_etal_1999,
barreiro_etal_2004,stolyarov_etal_2005}. Blind methods exploit the statistical 
independence between background and foregrounds. Among them, the Independent 
Component Analysis \citep[ICA, see][ and references therein]{amari_chichocki_1998}
has provided interesting results on simulated 
data in total intensity \citep{baccigalupi_etal_2000,maino_etal_2002} and polarization 
\citep{baccigalupi_etal_2004,stivoli_etal_2006a}, as well as on the real data 
by \COBE~\citep{maino_etal_2003} and $BEAST$ \citep{donzelli_etal_2006}; these 
results are based on an implementation of the ICA technique capable to rapidly 
achieve convergence to the solution if the hypothesis of statistical 
independence among the signals is verified \citep[\fstica][]{hyvarinen_1999}. 
Recently, hybrids techniques merging the blind and non-blind categories 
and largely based on a parametric description of the foreground emission 
have been proposed \citep{delabrouille_etal_2003,patanchon_etal_2005,eriksen_etal_2006}. 

As we already mentioned, the design and testing of the component 
separation algorithms has been successful on real data. This, 
together with the results presented here, indicates that these techniques 
may become standard data analysis techniques, complementary 
to the methods of CMB extraction exploited so far. In this 
work we apply the \fstica~algorithm to the total intensity 
\WMAP~3yr data. Our purpose is to identify the CMB among the 
separated components and study its pattern in comparison with 
the findings of other works exploiting different 
techniques. The paper is organized as follows. In Section \ref{fastica} 
we outline the main features of the \fstica~algorithm. In Section 
\ref{sims} we evaluate its capabilities on simulated \WMAP~data. 
In Section \ref{wmap} we apply the \fstica~to the \WMAP~data 
and study the recovered CMB. Finally, in Section \ref{conclusions} 
we draw our conclusions.

\section{Component Separation Problem}
\label{fastica}

We report here a brief description of the \fstica~algorithm, 
focusing on how the data are modeled and its principal assumptions. 
For further details we refer to the original theoretical papers in 
signal processing \citep{hyvarinen_1999,hyvarinen_oja_2000} and to 
the CMB applications, in total intensity \citep{maino_etal_2002} and 
polarization \citep{baccigalupi_etal_2004}.

Let us suppose that the sky radiation at a given frequency $\nu$
is the superposition of $N$ different
physical processes and that frequency and spatial dependencies can
be factorized into two separated terms:
\beq
\tilde{x}(\vect{r},\nu) = \sum_{j=1}^N \bar{s}_{j}(\vect{r})f_{j}(\nu)\, .
\eeq
By definition, and without loss of generality, the $f$ functions may be 
defined to be $1$ at a given reference frequency. In this case the signals $s$ 
represent the actual emissions at that frequency.
In order to exploit the different spectral behavior of CMB and foreground 
emissions, an $M$-frequencies experiment is usually exploited. Individual 
detectors are also coupled with an optical system, whose beam pattern is in 
general modeled, at each frequency, as a position invariant point spread 
function (PSF) $B(\vect{r},\nu)$. 
In addition any real experiment adds instrumental noise $\epsilon_\nu(\vect{r})$
in the output signal. Indicating with $\nu_{1},...,\nu_{M}$ the $M$ 
frequencies of a given experiment, one can define the scaling coefficients 
$a_{ij}=f_{j}(\nu_{i})$, and construct the $M\times N$ mixing matrix {\bf A} 
accordingly. Within these assumptions the observed signal is expressed by 
\beq
{\bf{x}}(\vect{r}) = {\bf{A}} {\bar{\bf{s}}}(\vect{r}) * B(\vect{r}) +
{\bmath{\epsilon}}(\vect{r}) = {\bf{A}} {\bf{s}}(\vect{r}) + {\bmath{\epsilon}}(\vect{r})\, ,
\eeq
where for each position $r$, {\bf x} and {\bf $\epsilon$} 
are vectors with $M$ rows, and the star represents the convolution 
of the PSF with the sky signals $\bar{\bf s}$, indicated simply as 
${\bf s}$ afterward;
note that we further assumed that the beam function is 
frequency independent i.e. the signals at different frequencies 
are smoothed at the same angular resolution.

The problem is solved obtaining both the mixing matrix ${\bf{A}}$ 
and the signals ${\bf{s}}$ from the observed data
${\bf{x}}$. In the ICA approach, the lack of determination 
in the problem is compensated by assuming that the signals to recover 
are random realizations of independent distributions: this means that 
the joint probability distribution of the superposition is a product of 
the single ones for each signal; in particular this implies that all of them, 
but at most one, have non-Gaussian distributions. The actual separation 
is achieved by means of linear combinations of the input data at 
different frequencies, corresponding to the maxima of a suitable 
neg-entropy approximation and thus of the non-Gaussianity. The maxima 
are in the form of vectors ${\bf{w}}$, rows of the separation 
matrix ${\mathbf{W}}$ such that the transformed variables 
${\mathbf{y}}={\mathbf{W}}{\mathbf{x}}$ are indeed independent 
components. The neg-entropy approximation is achieved by a non-linear 
re-mapping of the input data \citep{hyvarinen_1999,hyvarinen_oja_2000} 
exploiting suitably functions $g$, i.e. $g(u) = u^3$, $g(u) = 
{\rm tanh}(u)$ and $g(u) = u {\rm exp}(-u^2)$ where $u$ are the
principal component projected data.
In the following these functions will be indicated as $p$, $t$ and $g$. 
It is not possible to decide a priori which function will work better 
on a given data set: this indeed depends on the statistics of the 
independent signals which are not known a priori. In particular $p$, 
which corresponds to the kurtosis, should be used for sub-Gaussian 
components but it is strongly sensitive to outliers in the distributions; 
$g$ is for super-Gaussian signals while $t$ is a general purpose function 
\citep{hyvarinen_1999}. Therefore this optimization step is something
that have to be verified on a case-by-case basis. 
Similarly to previous works, 
in the application to the \WMAP~data we have verified that best performances 
are obtained with $g$ and we present results obtained with this function only.\\
Once the separation matrix ${\bf{W}}$ is obtained the underlying components are 
given by ${\bf{x}} = {\bf{W}^{-1}}{\bf{y}}$.  
This equation allows us to derive the frequency scalings for each independent 
component \citep{maino_etal_2002}: 
the scaling between $\nu$ and $\nu '$ of the $j^{\rm th}$ component is given by 
the ratio of ${W}^{-1}_{\nu j}/{W}^{-1}_{\nu ' j}$. This ratio is directly related to 
the spectral index of the frequency spectrum of the component which, for a power law
behavior, is written as $\beta = {\rm log}[W^{-1}_{\nu j}/W^{-1}_{\nu'j}]/{\rm log}[\nu/\nu']$.\\
The signal-to-noise ratio for the reconstructed components can also 
be recovered, as the noise enters into the separation process through 
the noise covariance ${\bf{\Sigma}}$ \citep{hyvarinen_oja_2000}. Therefore, 
noise constrained  realization ${\bf{n_x}}$ for each frequency channels 
can be built and combined accordingly to the weights obtained by the 
separation matrix ${\bf{W}}$. This is exactly what we have done for
evaluating the noise angular power spectrum in the recovered CMB component, 
as it was done for the \fstica~application to the \COBE~and $BEAST$ data 
\citet{maino_etal_2003,donzelli_etal_2006}. 

\section{Calibration with simulations and ideal performances}
\label{sims}

In order to construct an expectation of the results we obtain when applying
\fstica~to the \WMAP~3yr data, we first tested the algorithm by means 
of Monte Carlo (MC) simulations. These give hints on performances in ideal
situation and could be used, when compared to results from real data,
to judge the accuracy of sky and instrument model adopted.

\subsection{Simulation pipeline}
We created synthetic skies at the 5 \WMAP~frequency bands 
(23, 33, 41, 61 and 94~GHz) including
instrumental noise and smoothed with a Gaussian beam
of 1 degree FWHM. The sky signal is made of the CMB and the three major diffuse 
foreground emissions, namely synchrotron, free-free and dust. Their contribution
has been derived using the following models. The CMB template varies in the MC 
chains, corresponding to Gaussian realizations of the theoretical no running best 
fit CMB angular power spectrum from 
\WMAP\footnote{http://lambda.gsfc.nasa.gov/data/map/powspec/wmap$_{-}$lcdm$_{-}$pl$_{-}$model$_{-}$yr1$_{-}$v1.txt}
obtained with the {\tt cmbfast} code \citep{seljak_zaldarriaga_1996}. 
We have verified that up to $\ell \lsim 400$, which are the
scales of interest for 1$^\circ$ smoothed data, differences between 3yr and 1yr best-fit
model are less than 1.5\%.
For synchrotron, we use the 
all sky template at 408 MHz by \citet{haslam_etal_1982}; note that this 
does not include the extension by \citet{giardino_etal_2002} on sub-degree 
angular scale in total intensity and polarization following the data in the 
radio band \citep{jonas_etal_1998,reich_reich_1986,duncan_etal_1999,uyaniker_etal_1999}. 
For the dust, the $100$ $\mu$m map by \citet{finkbeiner_etal_1999} has been 
used; the free-free template is derived assuming correlation with the 
H$_\alpha$ emission \citep{finkbeiner_2004}.
These templates have been scaled to \WMAP~frequencies assuming the weights 
given in Table~4 of the work by \citet{bennett_etal_2003b}. 
The sky maps are treated following the HEALPix sphere pixelization 
scheme\footnote{http://healpix.jpl.nasa.gov} at $N_{side}=512$ resolution
parameter corresponding to about 7 arcminutes pixels.
Our foreground model is based on 1-year \WMAP~
results. A more realistic approach to foreground emissions could use
either the K-Ka template for ``synchrotron'' emission or the
3-year MEM foreground results \citep{hinshaw_etal_2006}. However the use of
these maps introduces subtleties mainly related to the noise in the data. Although the K-Ka map
is likely a superior tracer of the synchrotron emission than Haslam
at \WMAP~frequencies, for our demonstration purposes the model used is
sufficient.

The complete pipeline considered is therefore based on the following steps:
\begin{enumerate}
\item simulating CMB sky according to the \WMAP~best-fit model convolved with
the channel-specific beams; 
\item adding foreground emission scaled to each frequency band using the 
\WMAP~weights specified above;
\item adding noise realizations according to the sensitivity maps provided by
the \WMAP~team; 
\item deconvolving the beam and smooth to 1$^\circ$ FWHM each frequency
channel simulation; 
\item applying \fstica~on the simulated set of maps. 
\end{enumerate}
From each simulation we collect the results building figures of merit 
both on the full-sky and excluding Galactic plane regions with the Kp2 mask, 
used by the \WMAP~team for their CMB analysis and described in 
\citet{bennett_etal_2003b}. 

\subsection{Simulations results and performances}

\begin{figure*}
\begin{centering}
\includegraphics[width=8cm]{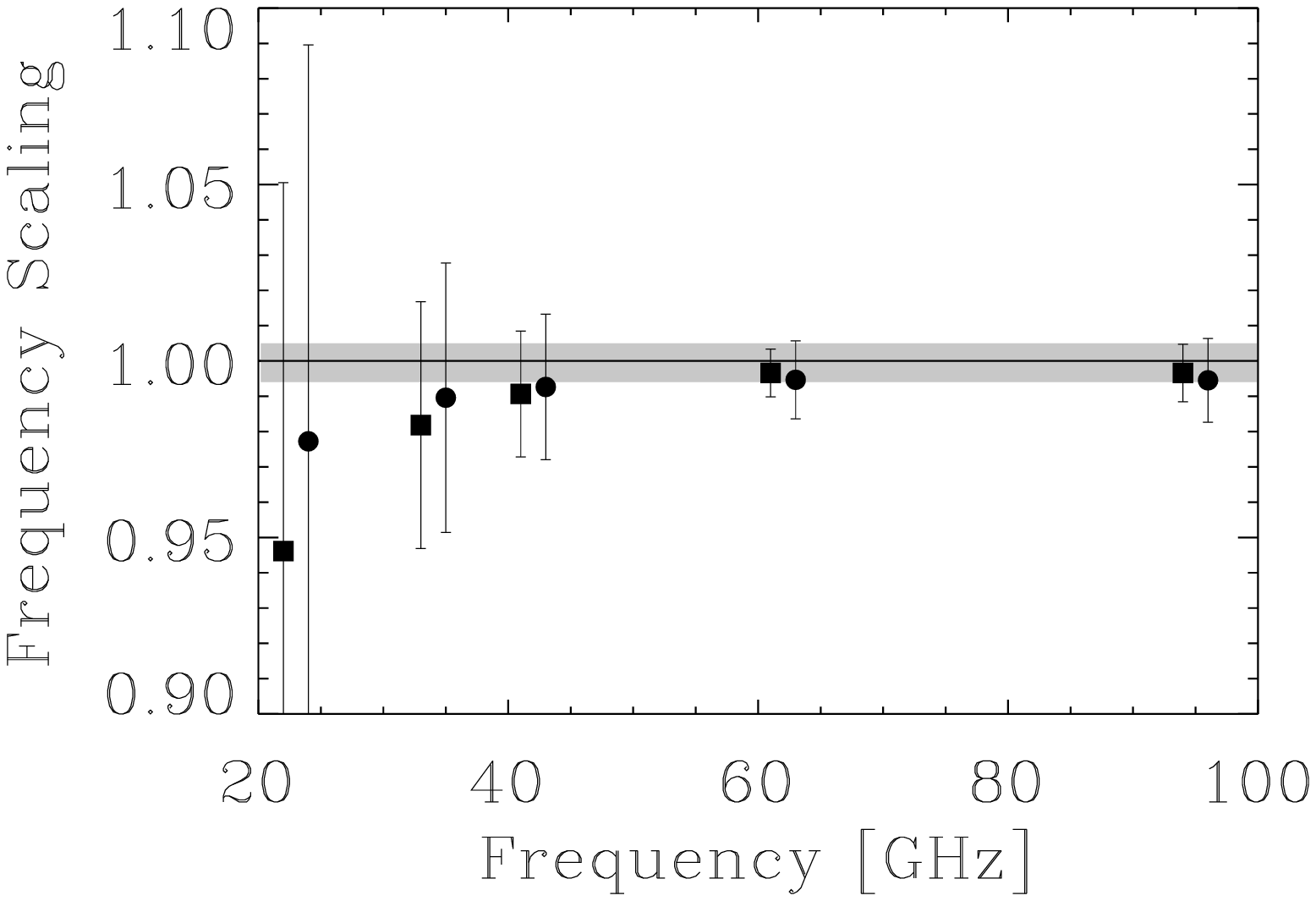}
\includegraphics[width=8cm]{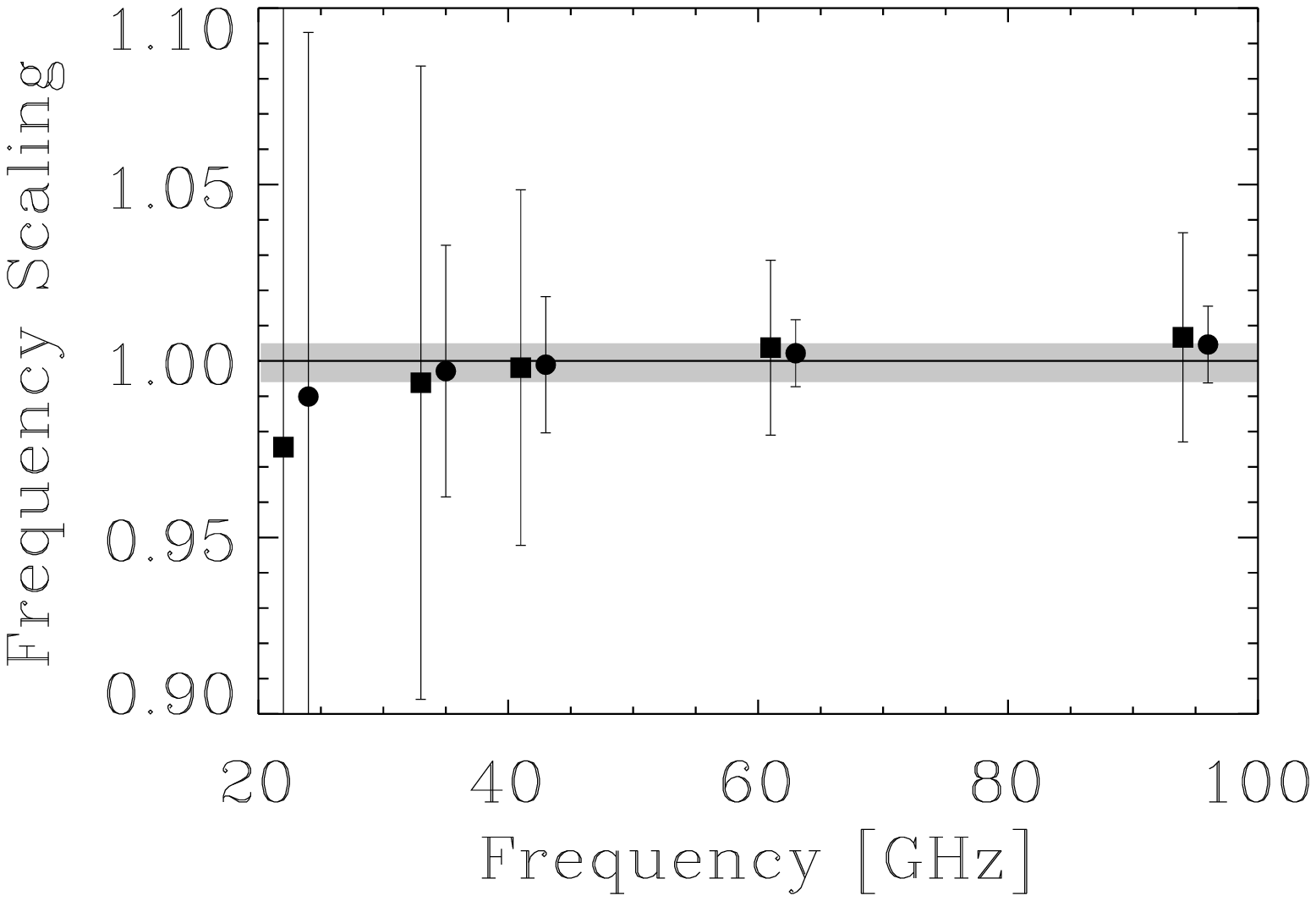}
\caption{ICA reconstructed CMB frequency scalings 
derived from 100 MC simulations. Boxes and circles
represent full sky and Kp2 scalings respectively.
The left (right) panel shows 
ideal results for noiseless (noisy) simulations. 
Also shown are the 
expected CMB frequency scaling and the 1-$\sigma$ error (gray band) from the 
\WMAP~analysis \citep{bennett_etal_2003a}.}
\label{sim_cmb_scal}
\end{centering}
\end{figure*}

\begin{figure*}
\begin{centering}
\includegraphics[width=5cm,angle=90]{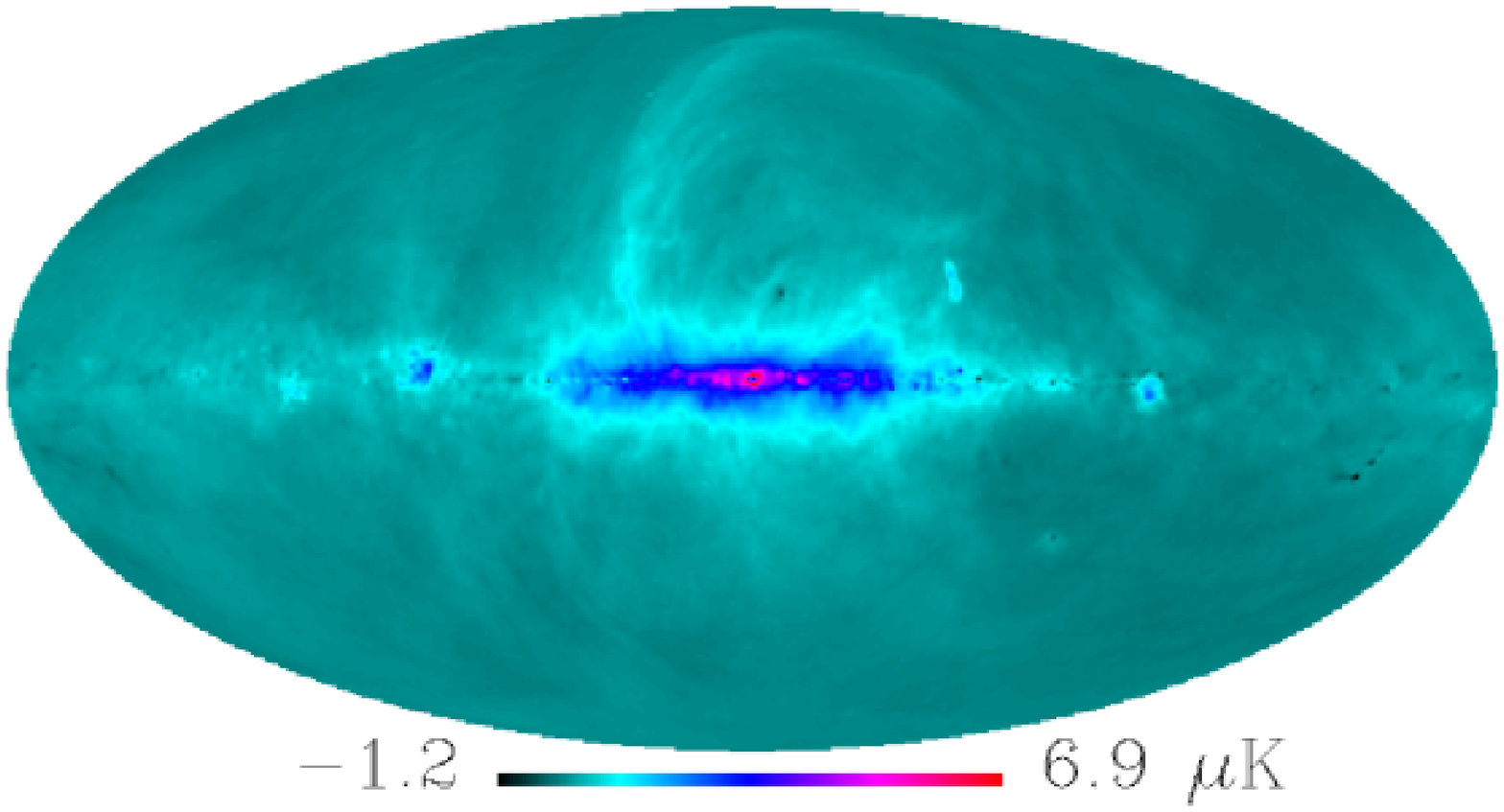}
\includegraphics[width=5cm,angle=90]{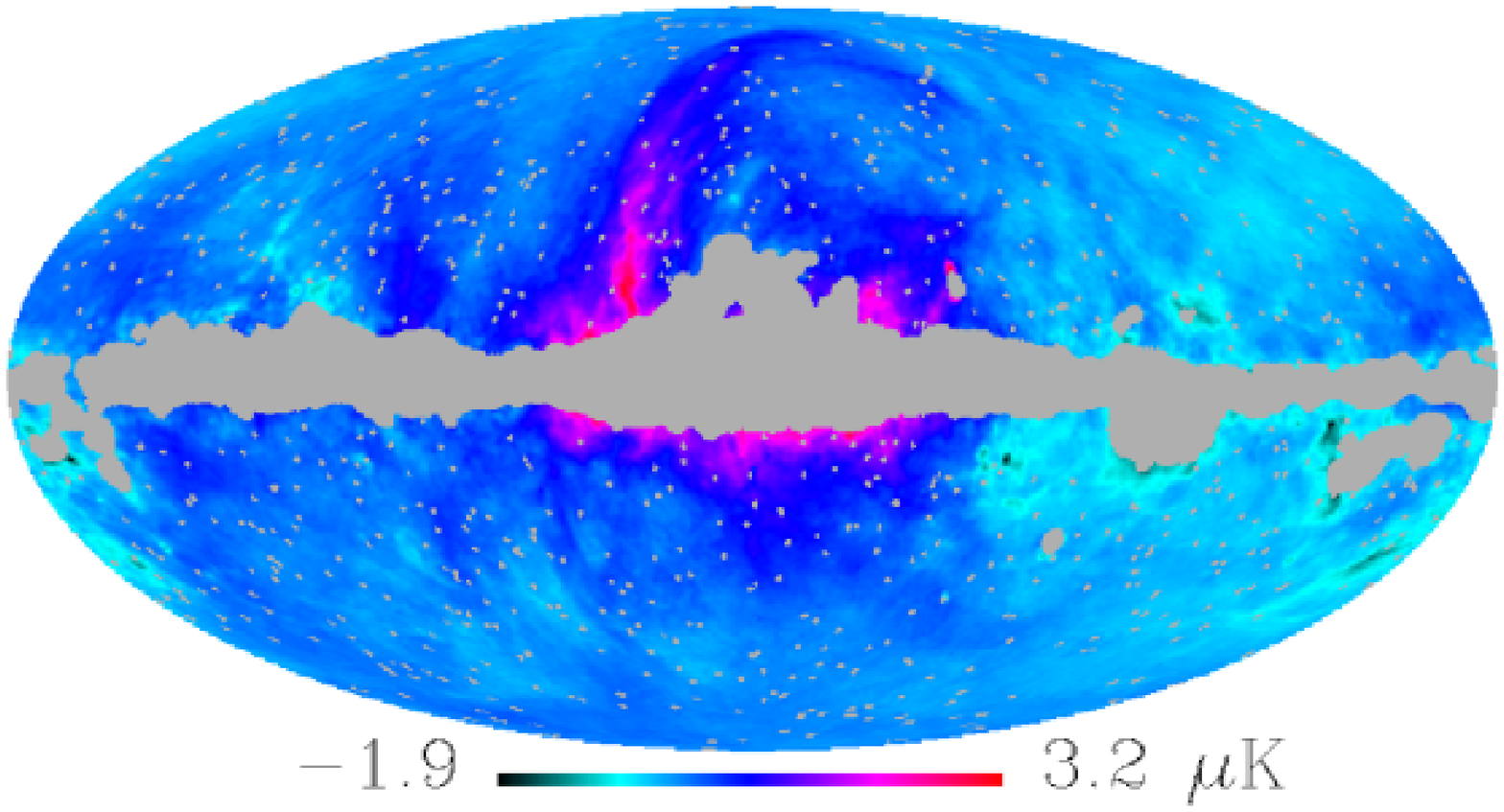}\\
\includegraphics[width=5cm,angle=90]{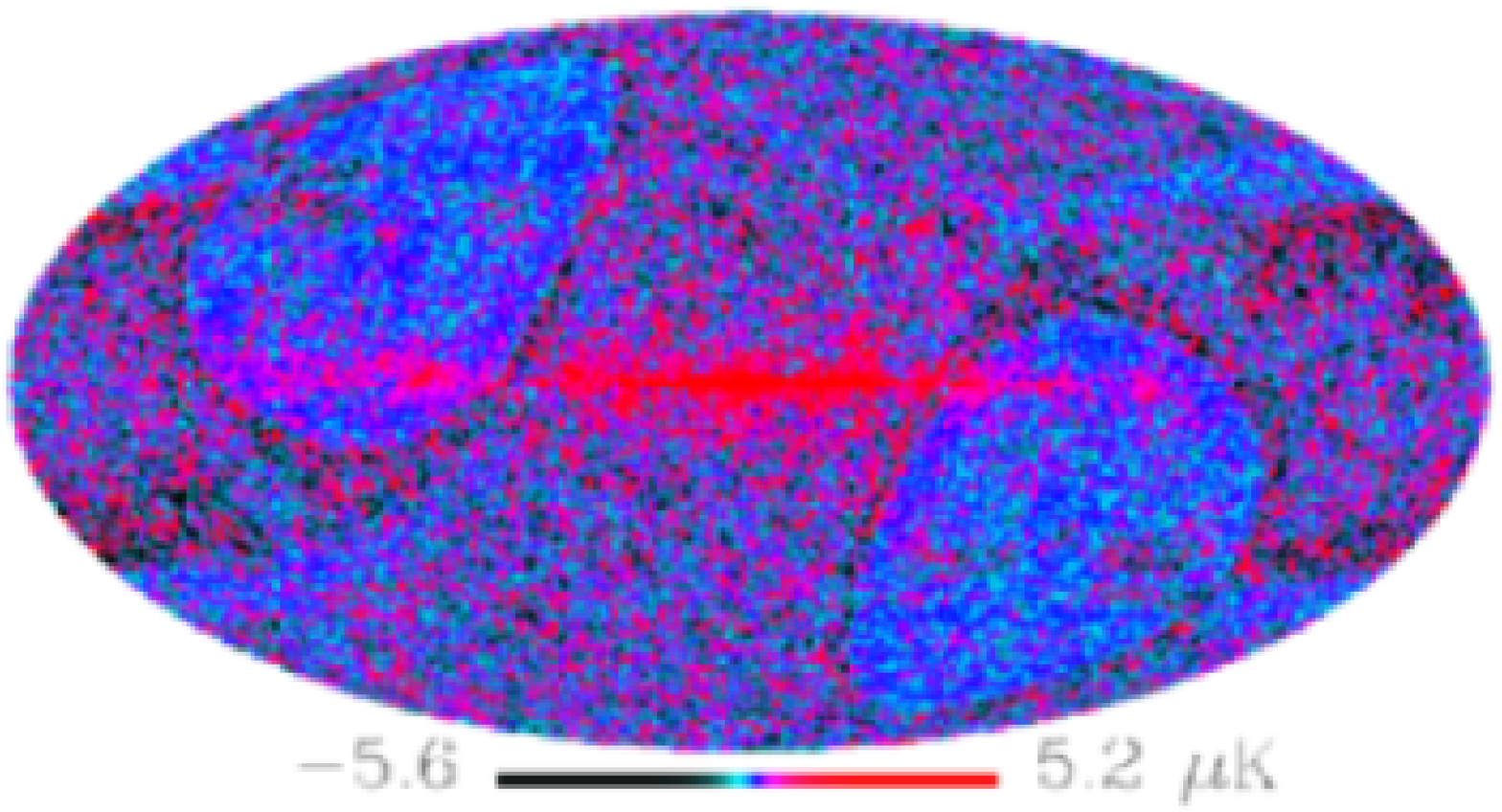}
\includegraphics[width=5cm,angle=90]{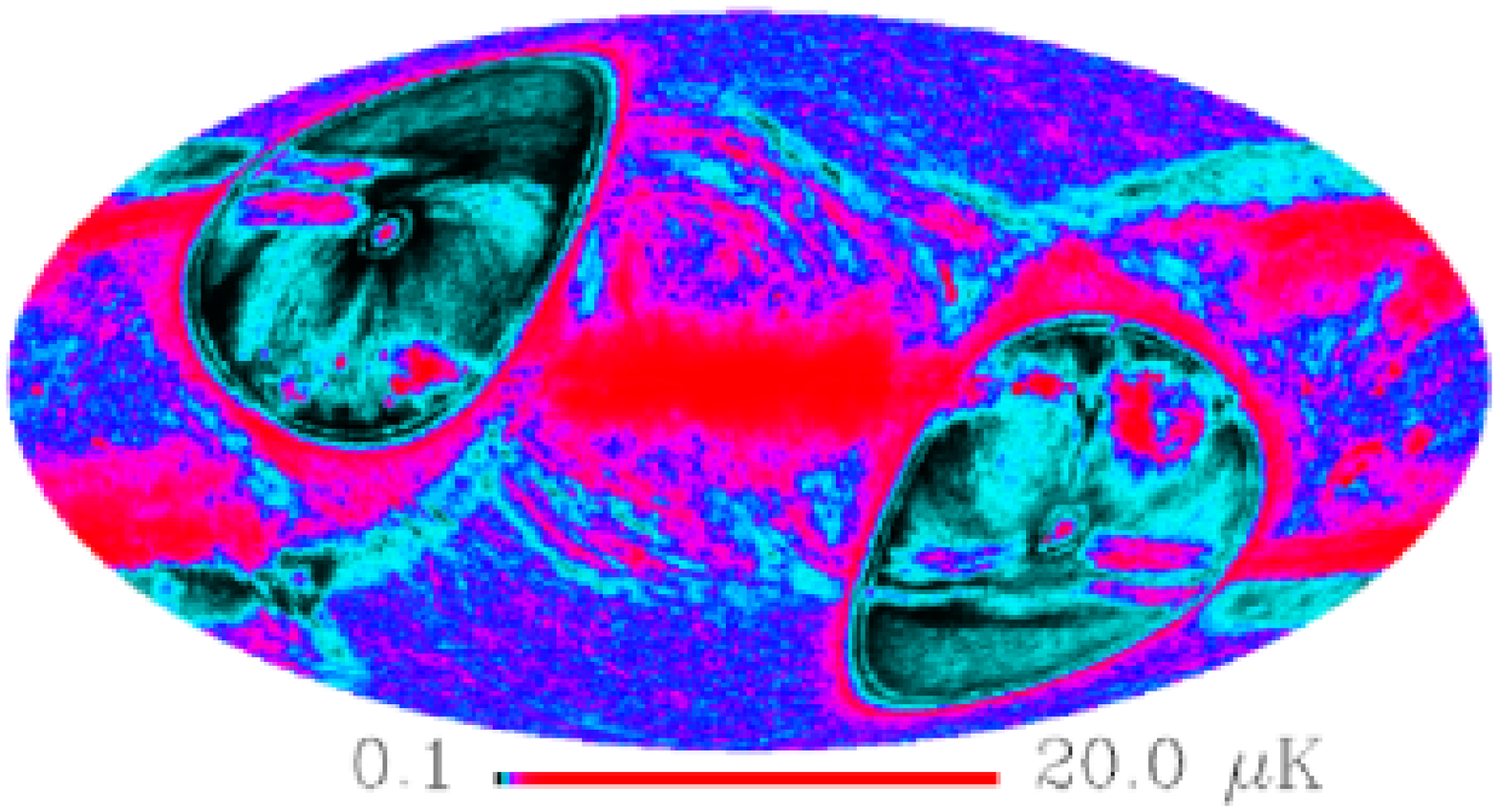}\\
\caption{Maps of the average and standard deviation of the difference between
reconstructed and input CMB maps. Each map is computed for a total of 100 simulations. 
The upper panels show the residual average for the ideal noiseless case when the reconstruction 
is done on the whole sky (left) and excluding the regions inside the Kp2 mask (right). 
The lower-left panel shows the residual average for the full-sky analysis with instrumental noise. 
The lower-right panel shows the standard deviation of the residual maps for full-sky and noisy case.}
\label{ave_std_maps}
\end{centering}\end{figure*}

From these set of simulations
we derived the distributions for the CMB frequency scaling 
in each frequency band. This is obtained by taking the sum of the elements in the row 
of the $\mathbf{W}$ matrix which corresponds to the extracted CMB signal and multiply 
this by the corresponding column in the $\mathbf{A}$ matrix. Anticipating what we have 
done on real data, we also considered different combinations of \WMAP~frequency 
bands with the aim of 
finding the optimal one for what concerns the CMB reconstruction. Such combinations are: 
QVW, KaQVW, KQVW, KQV, KaQV, KKaQV and KKaVW. Figure~\ref{sim_cmb_scal} shows results for the CMB
frequency scalings reconstructed by ICA for noiseless (left panel) and noisy (right panel) simulations for 
both the full-sky (filled box) and Kp2 (filled circles) analysis for the KKaQVW combination. 
The agreement with a pure black-body spectrum (thick horizontal line) is quite good and 
almost within 1-$\sigma$ limit of
\WMAP~calibration accuracy: low frequency channels (K and Ka bands) 
show larger uncertainties while if the bands 
considered are those where the CMB is stronger, the reconstruction is much more faithful.
The error bars in the figure are purely indicative of the expected performance 
of ICA on the real data by \WMAP. Indeed, the foregrounds are far too simple to 
accurately represent their real signal in the microwave band, at least 
for the low frequency components, synchrotron and free-free, where observations 
are limited in angular resolution, or indirect. 
Other unknowns are represented for example by the spatial variation of the frequency 
spectral index. Nevertheless the shown error bars give a 
flavor of the kind of accuracy of the results we show in the next Section. 
In addition, since for simulations we know exactly the input CMB sky, we evaluate 
both the average and the $rms$ of the residual map obtained subtracting 
the CMB input and output. We also consider noiseless simulations in order 
to check the impact of the instrumental noise on the quality of the reconstruction. 
In the upper left panel of Figure~\ref{ave_std_maps} 
we show the average residual map from 100 simulations using all \WMAP~channels in 
the ideal noiseless case: \fstica~removes the foregrounds to the level indicated in the 
figure, in units of thermodynamical temperature, with the strongest residual signal 
along the plane where the Galaxy is brighter. This is indeed expected since along the 
plane foreground emissions are expected to be correlated, violating one of the 
\fstica~assumptions. The peak-to-peak amplitude of the residual is around 
8$\mu$K. Excluding Galactic plane regions and point sources with the
Kp2 mask reduces sensibly the contamination (upper-right panel) down to $\sim 5\mu$K. 
Adding instrumental noise (lower-left) slightly decreases the quality of the 
reconstruction: the residuals are still present along the galactic plane and 
structures related to the actual noise
distribution in the sky appear. Although quite similar between the 
\WMAP~channels, the non-uniform noise distribution is interpreted by the algorithm 
as an additional ``signal'' component. However the peak-to-peak residual 
remains quite small being around $10.8\mu$K. Finally in the lower-right panel we 
show the standard deviation of the residual map: the largest deviations are again along 
the Galactic plane where a very bright spot is clearly visible at a level 
of $20\mu$K. However for the Kp2 analysis the overall rms decreases to less than 6~$\mu$K 
indicating a good quality in the \fstica~reconstruction; the scanning pattern of 
\WMAP~is also evident. \\
Simulations here include only four signal components to be extracted with 
five frequency channels, and in the actual implementation
the \fstica~tries to recover a number of components equal to the one of 
the frequency channels considered; also we do not know a priori the
order in which components are extracted. Adopting a criterion already 
used in previous works \citep{maino_etal_2003}, we were able to verify that 
the fifth component is clearly not physical, evaluating its 
signal-to-noise ratio; however its small amplitude 
slightly contaminates the other extracted components, and the best configuration 
for \fstica~is with four frequency channels at least in simulations. The
situation is slightly different in the application to \WMAP~real data due to the
physical properties of real foregrounds.

\section{Application to \WMAP~3yr data}
\label{wmap}

In this Section we show the results obtained by applying the 
\fstica~algorithm to the \WMAP~data. There are three main findings 
and criteria, namely the combination of the different \WMAP~channels, 
the results for the power spectrum and the distribution of the CMB 
power as recovered by the \fstica~across the sky, which are the subjects 
of the following sub-sections.

\subsection{Selection of input maps}
\label{maps_scalings}

\fstica~operates onto the \WMAP~3yr data smoothed at 1 degree FWHM 
at $N_{\rm side} = 512$, on the full sky and using the Kp2 mask. 
It is interesting to study the results for several combinations of the 
\WMAP~channels, selecting in particular the optimal one for what concerns 
the CMB reconstruction. We have performed our analysis considering the whole set of 
\WMAP~data as well as suitable subsets thereof. In particular we have set the 
minimum number of input channels to three taking the following combinations:
QVW, KQVW, KaQVW, KQV, KaQV, KKaQV and KKaVW. The first three include high frequency
CMB dominated channels: the rationale of adding to QVW the data in K or 
in Ka band is to provide the \fstica~dataset with a frequency band where 
the low frequency foregrounds dominate to better assess and remove the 
corresponding contamination to the CMB, and to check 
which between the K and Ka bands is more 
representative of such contamination. The next three sets exclude the W band data
in order to see which channel dominates the CMB reconstruction,
while the final one is included since the final \WMAP~3-yr CMB angular power 
spectrum for
$\ell > 12$ has been obtained by combining V and W bands data after subtraction
of the K-Ka map plus free-free and dust templates.
%
Several things have to be noted. 

\begin{figure*}
\begin{centering}
\includegraphics[width=5cm,angle=90]{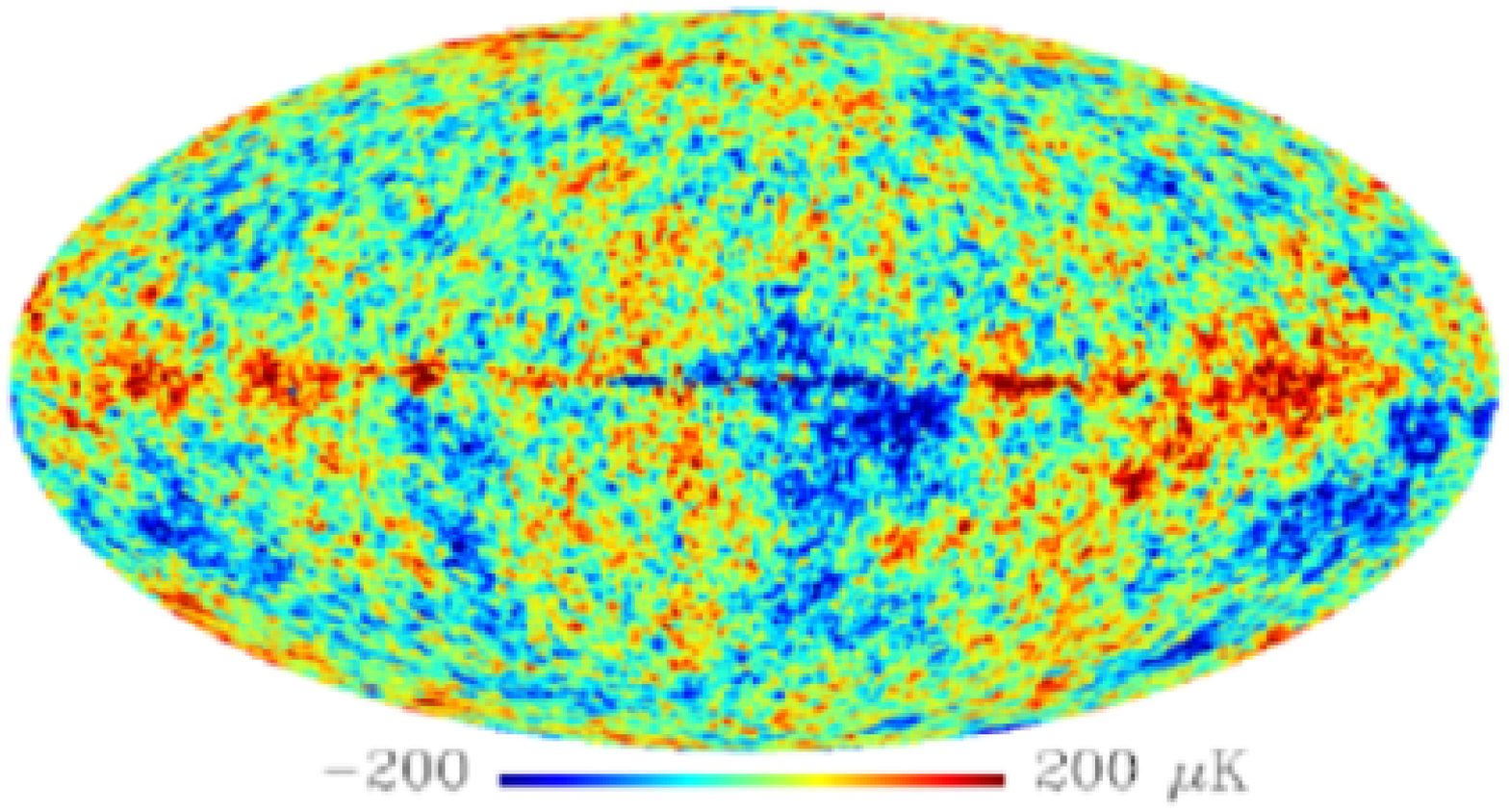}
\includegraphics[width=5cm,angle=90]{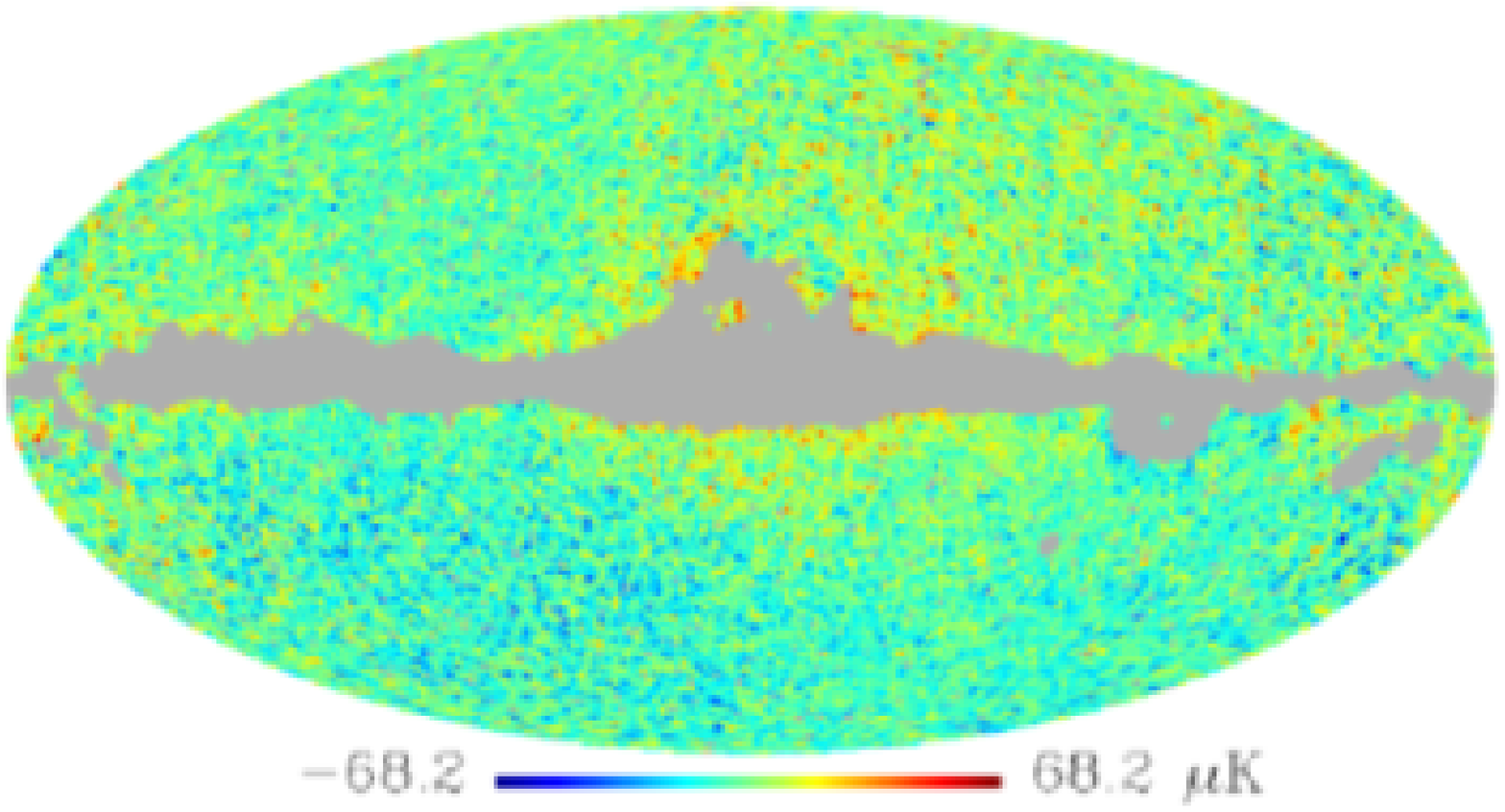}
\caption{Left - CMB ICA map for full sky KKaQVW analysis. The residual 
along the Galactic plane is evident. Right - Difference between full sky and Kp2 ICA CMB
maps: some residual Galactic signals is present near the edge of the sky mask and around
some bright sources. High latitudes, large scale residual patterns are also present.}
\label{wmap_ica_maps}
\end{centering}
\end{figure*}

First all the recovered full-sky CMB maps show clearly a residual
contamination along the galactic plane; we report in Fig.~\ref{wmap_ica_maps}
the \fstica~CMB reconstruction from the KKaQVW combination as well as the 
difference between this map and the corresponding one from the Kp2 analysis. 
\begin{table}
\begin{center}
\caption{{\sc FASTICA} weights for different channels combination}
\label{weights}
\begin{tabular}{ccccc}
\hline
\ K & Ka & Q & V & W \\
\hline
\multicolumn{5}{c}{\bf Full-sky}\\
\hline
\  0.001610   &  0.039723  & -0.932632 & 2.02027 & -0.128972 \\
\ -           & -          & -0.832358 & 1.93718 & -0.104825 \\
\  0.002866   & -          & -0.849484 & 1.94972 & -0.102067 \\
\ -           &  0.007923  & -0.947255 & 1.93514 & -0.095834 \\
\ -0.026772   &  0.104729  & -0.872566 & 1.79461 & - \\
\ -0.008567   & -          & -0.755391 & 1.76396 & - \\
\ -           & -0.034631  & -0.730927 & 1.76556 & - \\
\  0.108010   & -0.680878  & -         & 1.64439 & -0.071519 \\
\hline
& & & & \\
\multicolumn{5}{c}{\bf Kp2 sky cut}\\
\hline
\ -0.183053  & 0.248303   & -0.158422 &  0.767720 &  0.325453 \\ 
\ -          & -          & -0.810260 &  2.231690 & -0.421430 \\
\ -0.148842  & -          &  0.232536 &  0.748460 &  0.167486 \\
\ -          & -0.573262  &  0.524846 &  1.23660  & -0.188187 \\
\  0.074617  & -0.413025  & -0.327681 &  1.66609  & - \\
\  0.001962  & -          & -0.750511 &  1.74855  & - \\
\ -          & -0.436905  &  0.295693 &  1.14121  & - \\
\ -0.508637  &  1.18968   &  -        & -0.703798 & 1.02276 \\
\hline\hline
\end{tabular}
\end{center}
\end{table}

\begin{figure}
\begin{centering}
\includegraphics[width=8cm]{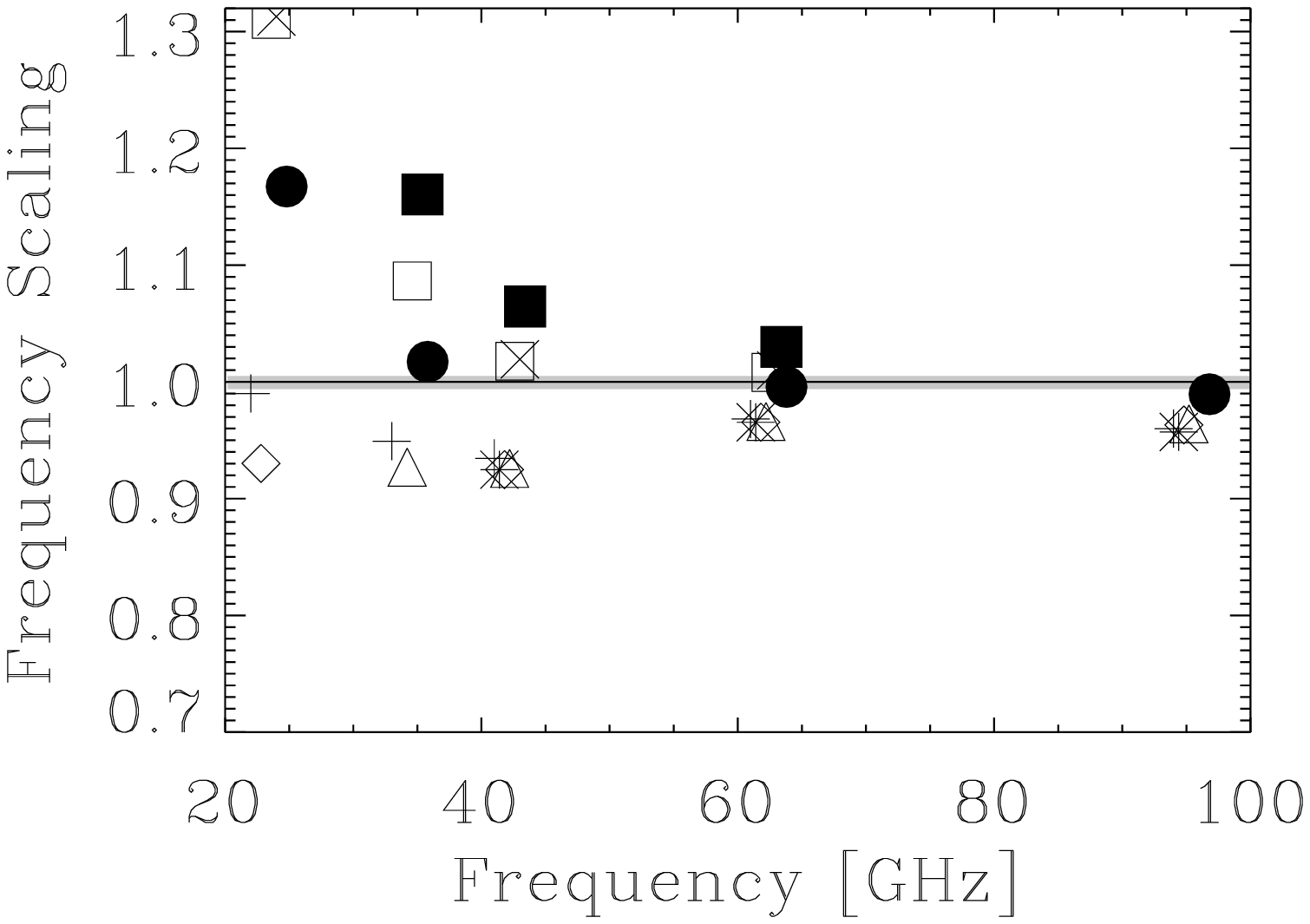}
\includegraphics[width=8cm]{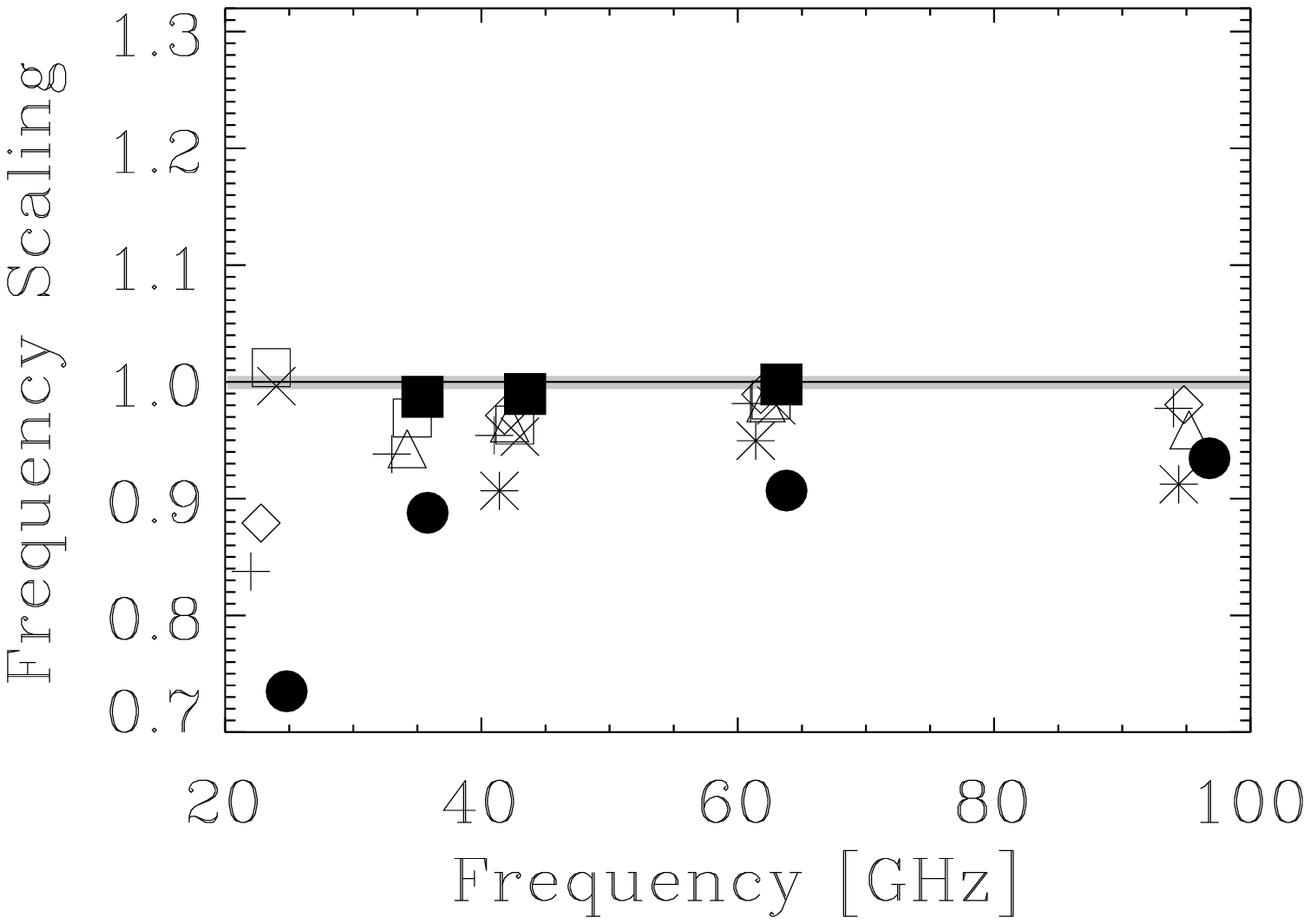}
\caption{CMB frequency scaling for full-sky (left panel) and
Kp2 analysis (right panel). Plus signs refer to KKaQVW combination,
asterisks to QVW, rombs to KQVW, triangles to KaQVW, empty square to KKaQV,
crosses to KQV, filled squares to KaQV and filled circles to KKaVW.}
\label{wmap_cmb_scalings}
\end{centering}
\end{figure} 
The Galactic residual contamination 
is due to the fact that along the plane the foreground emissions are expected 
to be correlated, violating one of the \fstica~assumptions, as it was already 
evident in the simulations in the previous Section, although in that case the effect 
is much smaller in amplitude. 
This is an indication of the fact that our sky model is too far simple to
properly reproduce the case with real data. This issue could be solved 
in principle by using a sort of Internal Template approach as
exploited by \citet{hansen_etal_2006} and currently under study 
\citep{stivoli_etal_2006b}. In the difference map large
residuals are mainly located near the edge of the sky mask and around 
bright point sources: this is not surprising as we will see shortly.
Moreover large scale residual is also present and we will comment
on this when considering power spectrum analysis.

In addition we report in Fig.~\ref{wmap_cmb_scalings} \fstica~reconstructed
CMB frequency scalings for both full-sky (left panel) and Kp2 sky cut (right panel)
analysis. The improvement in the reconstruction when working with the Kp2 mask
is evident at least for the low frequencies (K, Ka and Q bands) but not marked.
Indeed the two CMB dominated channels are closer to the expected scaling
for the full sky analysis than when Kp2 mask is applied: 
deviations 
are clear for
the QVW combination and particularly for 
KKaVW channels which are quite peculiar as we will see below.
This indicates 
that these three channels (QVW) do not provide enough information to \fstica~
on the low-frequency contamination to properly perform a CMB cleaning. 

For what concerns the frequency scaling the optimal combination is the one with KQVW
where the deviation from the expected value of 1 for the highest frequency channels
is smaller. This is an interesting result: it indicates that according to the 
figure of merit adopted, the low foreground emission is better ``read" by \fstica~
by means of the information in the K-band which is then subtracted optimally 
in the Q, V and W ones. On the other hand, this maximum in the quality of the 
CMB reconstruction lies in a plateau, in the sense 
that the results for other combinations are stable and pretty consistent with 
each other, as the rest of the analysis shows.

In Table~\ref{weights} we also report the \fstica~weights for the different 
channel combinations and sky cuts. It may be immediately noted that the 
number variation is macroscopic, even within a given case, either on the 
full sky or outside the Kp2 mask. Indeed, a different combination of the 
\WMAP~channels implies a different relative distribution of the background 
and foreground components, making \fstica~converging to a different 
separation matrix. The difference in the dataset by selecting different 
channel combinations is enhanced by the fact that \WMAP~foregrounds are 
real, e.g. they possess a space varying spectral index. On 
different combinations \fstica~deals with a different sky rather than 
the same one at different frequencies, making the problem and the solution 
markedly varying. Despite of this, the reconstructed CMB power spectrum is 
pretty consistent in the different cases, as we see in a moment, and some
general trends are clearly visible. First of all the dominant component 
with respect to the weight amplitude is always V-band for both full sky and Kp2 
analysis. This is not surprising since it has the lowest foreground 
contamination. Furthermore in full sky analysis Q-band has always large 
negative weights almost stable with respect to the different combinations and
similarly to the W-band they are both stable and negative.
The Kp2 analysis assigns again largest weights to the V-band, apart from 
the KKaVW combination, while 
the overall behaviour of the results is less stable possibly due to the
difference in the sky that \fstica~has to deal in each case, due to
the fact that the data are real in the present case.
A possible trend is that
when both K- and W-bands are included, the weights in K are positive while
they are negative for W-band. This could be related to the fact that the
W band traces dust possibly both thermal and anomalous 
component, and K presumably contains some anomalous dust
too.
We also verified that the same instability in the weights recovered by the 
\fstica~occurs considering the other masks provided by the \WMAP~team
(namely the Kp0 and the Kp2 extended mask). 
Specifically, keeping the channel combination fixed but varying the 
mask, the weights variation has roughly the same magnitude as in 
Table~\ref{weights}. The interpretation we give is the relevance of the 
foreground signal on the ridge of a given mask, where the foreground 
themselves are more intense. Changing the cut is equivalent to include 
or exclude areas where the foreground emission is intense, and 
most likely exhibits different properties, as expected for realistic 
astrophysical emissions. Thus \fstica~deals in each case with a 
different problem, with the result of finding different weights of the 
different channels which maximize the independence of background and 
foregrounds. This situation is clearly evident, as reported
commenting the derived frequency scalings, for the
KKaVW combination: W-band is almost left untouched while
other frequencies combine to create the ``foreground'' map to be 
subtracted from the W-band data. This is completely different
from the previous findings and it is another indication of what
reported before: different combinations are indeed different skies
to be analysed. In this specific case without Q-band and without
the region near the galactic plane, \fstica~detects different
signal statistics as demonstrated by the different weights obtained.
However the quality of the CMB reconstruction is still
quite good given the high statistical independence of CMB
with respect foregrounds as reported in the following analysis
on the power spectrum.

\begin{figure*}
\begin{centering}
\includegraphics[width=12cm]{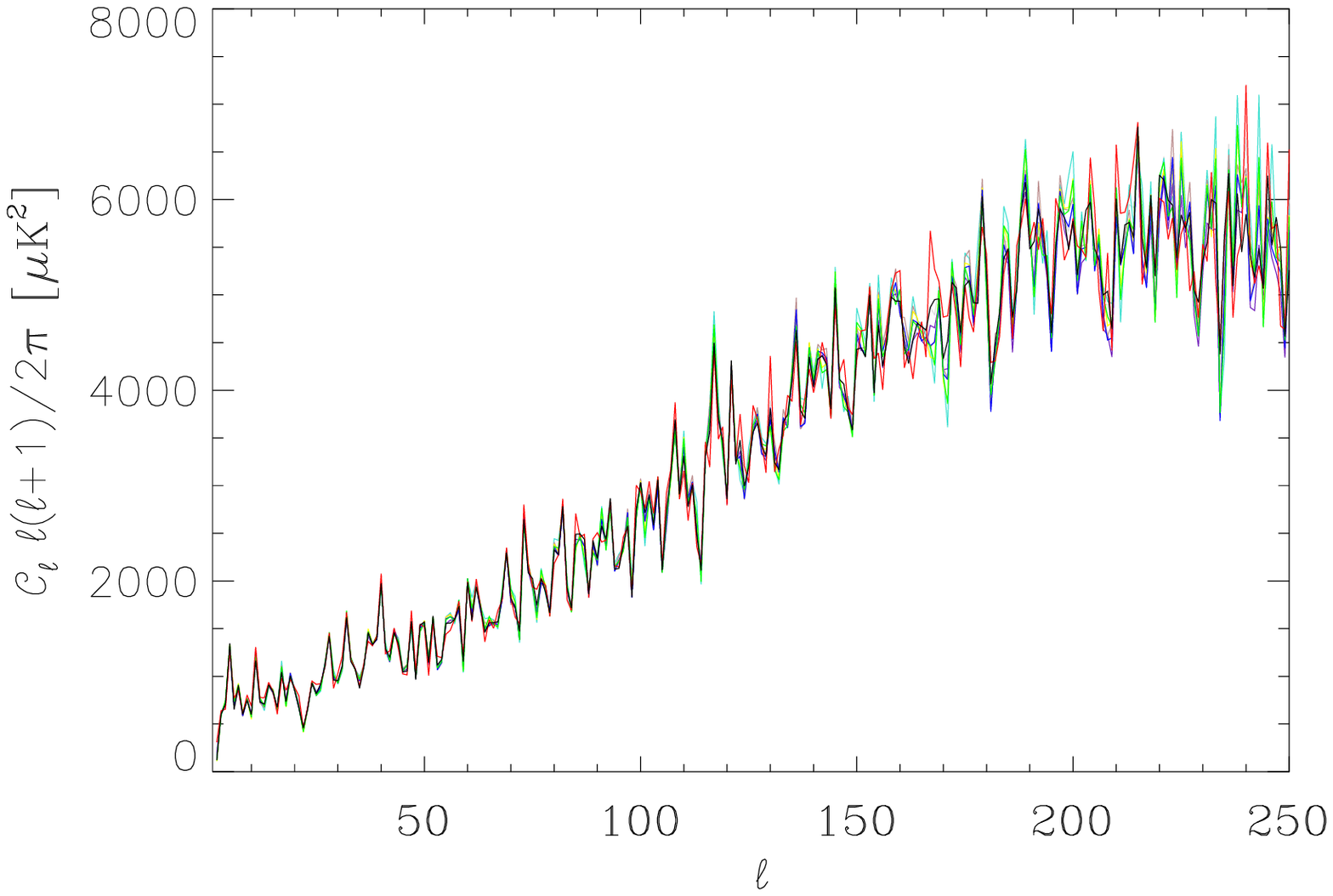}
\includegraphics[width=12cm]{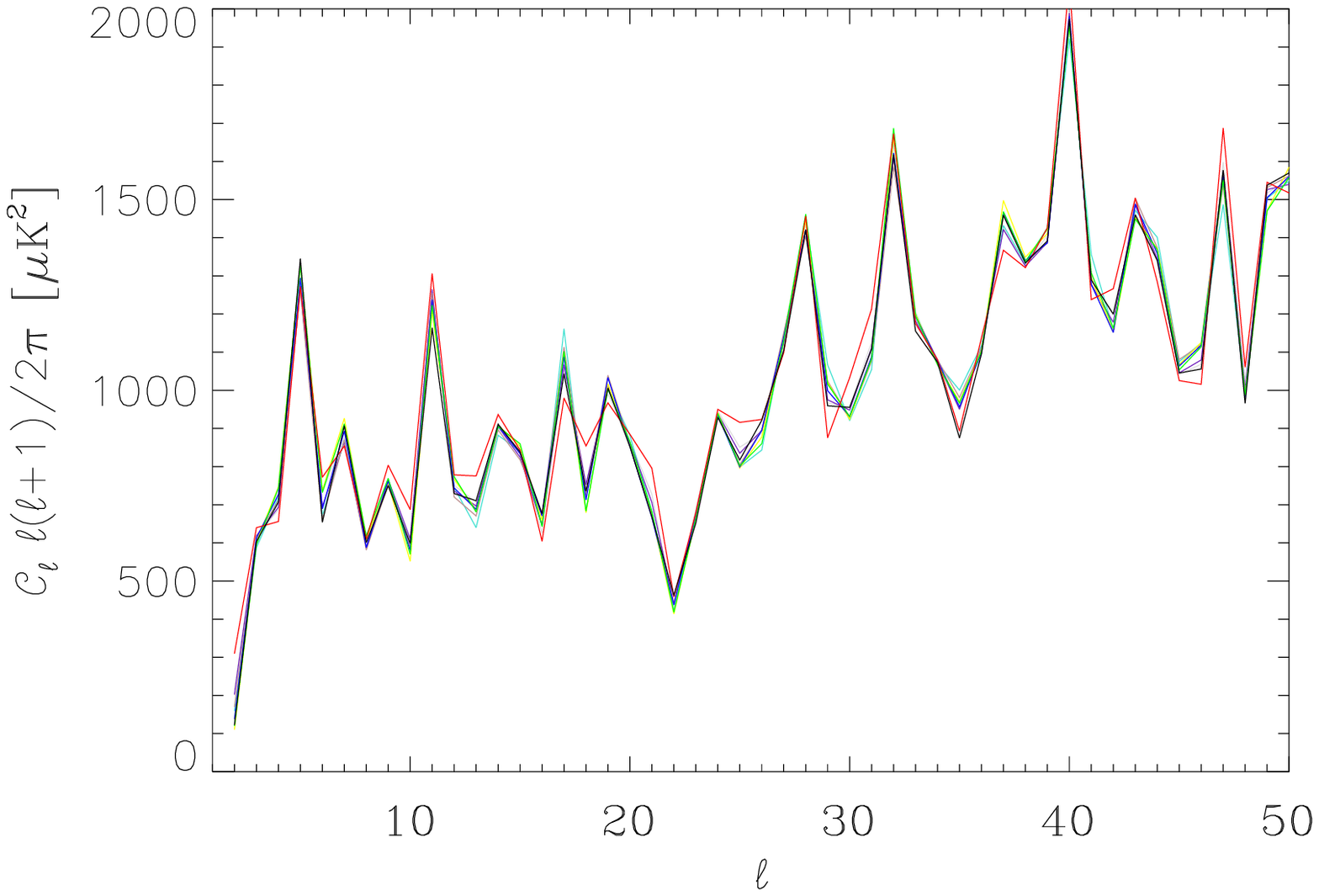}
\caption{The power spectrum from all the \WMAP~channel combinations 
on Kp2 compared with the full MASTER \WMAP~3yr power spectrum (black thick
solid line). The top panel shows the analysis considering 
all multipoles, while the bottom panel reports results for low $\ell$s 
only, see text. Channel combinations are color coded: 
red for KKaQVW, green for QVW, blue for KQVW, yellow for KaQVW,
brown for KKaQV, indigo for KaQV, turquoise for KQV and grey for
KKaVW.}
\label{cl_comparison}
\end{centering}
\end{figure*}

\subsection{Power Spectrum}
\label{ps}

Here we derive the angular power spectrum from the ICA CMB maps 
discussed above using the MASTER algorithm \citep{hivon_etal_2002}, 
restricting to the cases in which the Kp2 mask has been used. 
This means that we also consider
the full sky results after applying the Kp2 mask.
Noise biases are obtained from noise-only simulations of the 
\WMAP~frequency channels, combined with derived 
ICA weights (see Table~\ref{weights}) specific of each channel combination considered.\\
The final \WMAP~3yr power spectrum have been estimated using a combination of a 
maximum-likelihood approach for multipoles $\ell < 12$ and a MASTER technique
for higher multipoles. However since we adopt a MASTER approach also at low multipoles
for a proper comparison we report here the full MASTER \WMAP~3yr power spectrum
kindly provided by the \WMAP~team \citep{hinshaw_private_2006}.
Figure~\ref{cl_comparison} presents our results for the selected channel combinations 
compared to the \WMAP~3yr CMB power spectrum. 

There is a complete consistency between the power spectra obtained from 
different channel combinations and the \WMAP~3yr results. 
The analysis on the low $\ell$ part of the spectrum shows again a general good agreement
between ICA results and \WMAP~3yr data almost regardless of the channel 
combination considered. 
This is interesting since the \WMAP~result has been obtained considering 
only high frequency channels after subtraction of foreground template as traced by the
difference between low frequency channels. This result represents an indication
of the high quality results that \fstica~can obtain in terms of CMB reconstruction
on real CMB data.

\begin{figure}
\begin{centering}
\includegraphics[width=12cm]{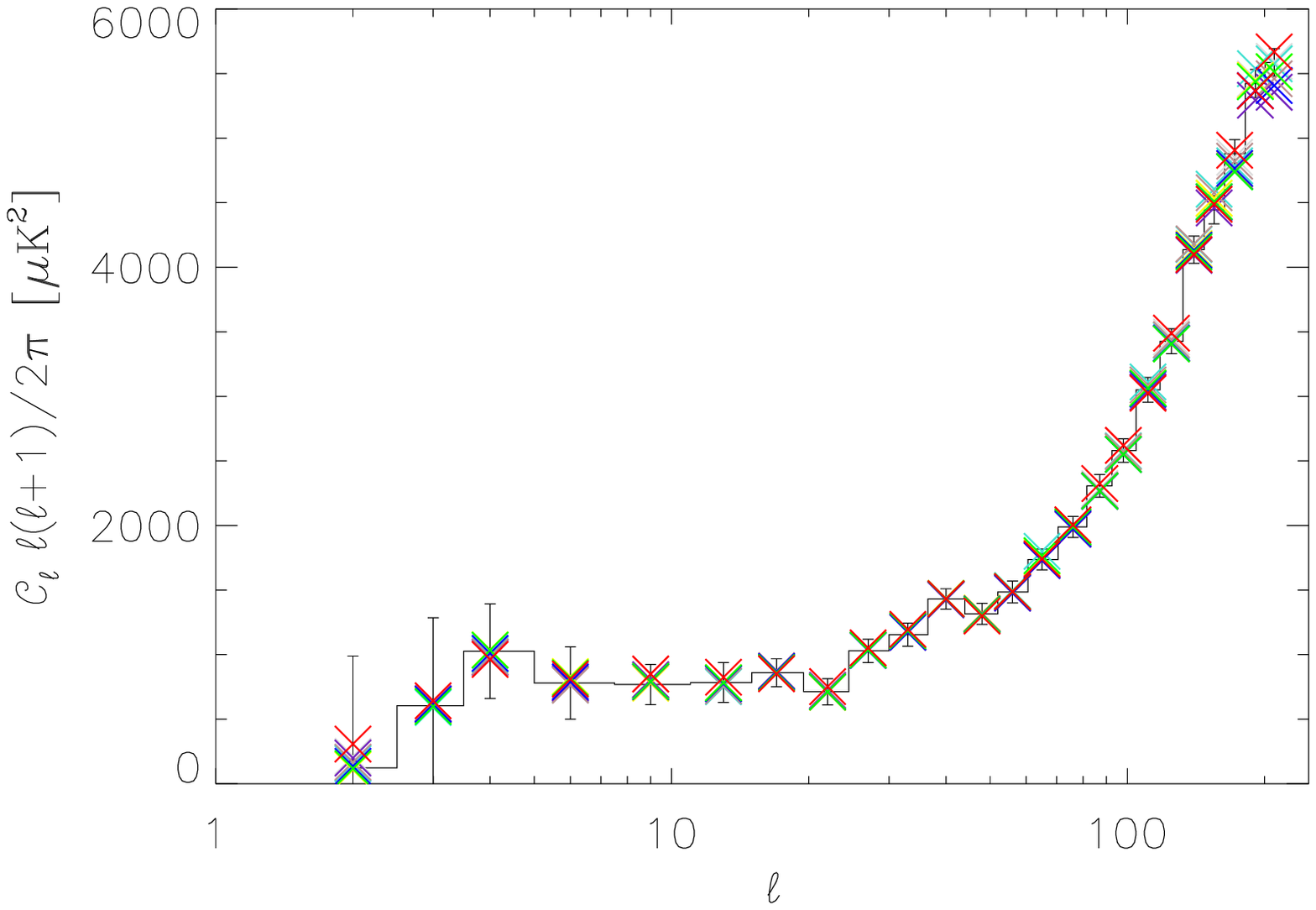}\\
\includegraphics[width=12cm]{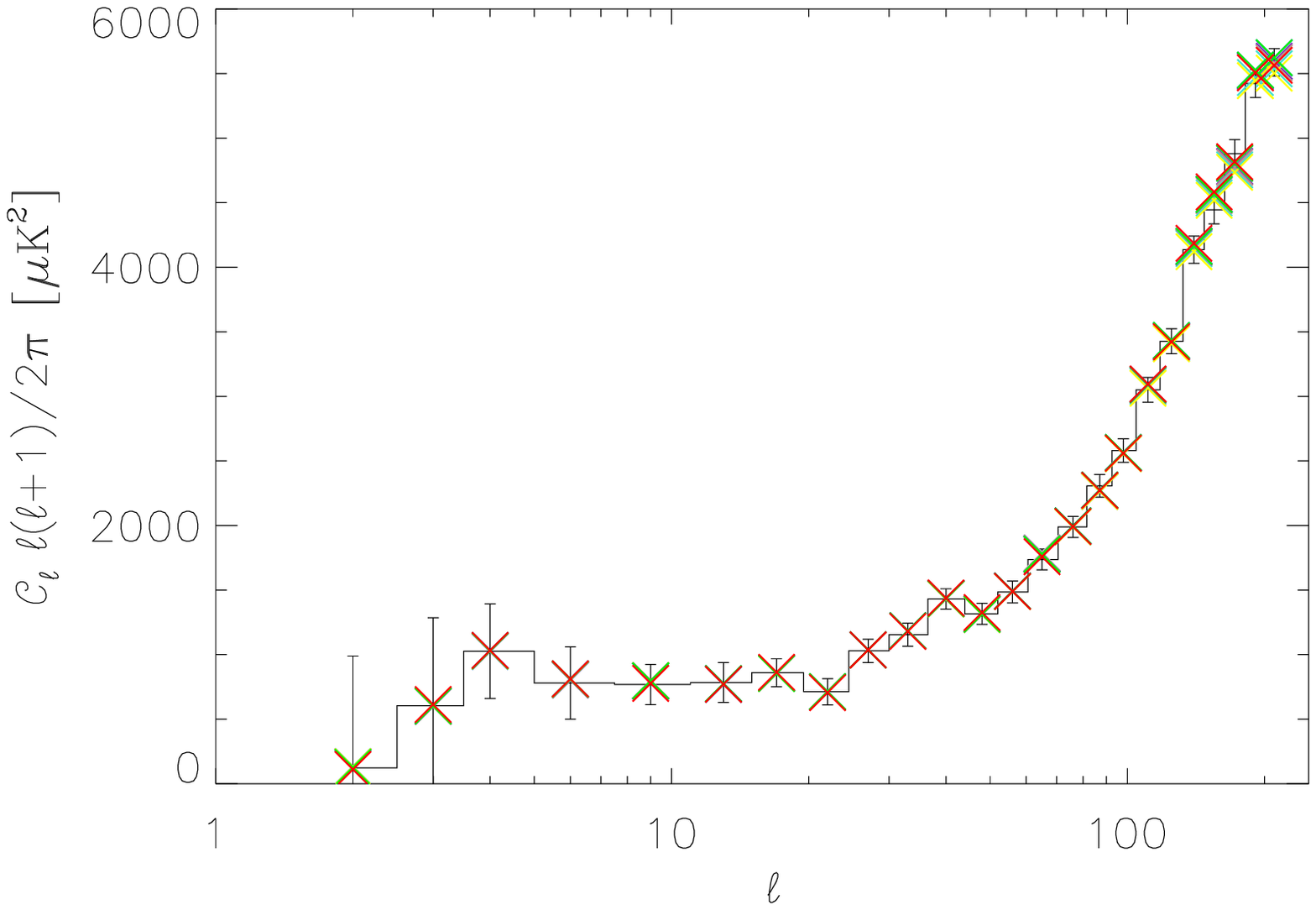}
\caption{The binned power spectra from all \WMAP~channel combinations compared
with the full MASTER \WMAP~3yr power spectrum for a complete Kp2 analysis
(upper panel) and full sky analysis with Kp2 mask applied when computing
the power spectrum (lower panel). Channel combinations are color coded
as in the previous figure.}
\label{cl_binned}
\end{centering}
\end{figure}

We report in Figure~\ref{cl_binned} (upper panel) a similar comparison but for the binned power spectra 
and the agreement is evident both on the low multipoles as well as to
higher ones up to $\ell \simeq 150$. Large spread in the results is present for 
the highest bins. Adopting the consistency with the \WMAP~3yr power spectrum 
as a figure of merit, we can 
judge which is the optimal combination: this is once again the KQVW, indicating that
the best tracer of low frequency foreground contamination at high frequencies is 
represented by the K band data.
On the bottom panel we show similar results from the full sky analysis when the Kp2
mask has been applied: the agreement with \WMAP~results are even more evident for
both low and high $\ell$. This is an indication of the fact that even in presence of
strong and possibly correlated foreground on the galactic plane, \fstica~is not only 
still able to properly recover the CMB pattern at high galactic latitudes but it performs
better than in the case of a pure Kp2 analysis. This means that the level of signal correlation
along the galactic plane, that violate one of the ICA assumptions, does not compromise
the reconstruction at high galactic latitudes and that the regions near
the plane included in the full sky analysis are useful for better distinguish different
signal statistics.

\subsection{Power Asymmetries}
\label{pow_asym}

Given the quality of the results obtained so far especially when considering 
regions outside the Kp2 cut, we push the analysis further. 
Several authors \citep{eriksen_etal_2004b,hansen_etal_2004,hansen_etal_2006},
reported an unevenly distribution of large-scale power in the \WMAP~1yr 
data. The asymmetry is maximized in the direction defined by a north pole at $(\theta,\phi)=(80^{\circ},57^{\circ})$ 
(Galactic co-latitude and longitude). Different and independent techniques found
that in such reference frame the southern hemisphere has significantly more
power than the northern hemisphere for $\ell<40$. These findings
have been confirmed in the \WMAP~3yr data. We therefore investigate
the ICA CMB maps searching for this asymmetry.

\begin{figure*}
\begin{centering}
\includegraphics[width=6.3cm,angle=90]{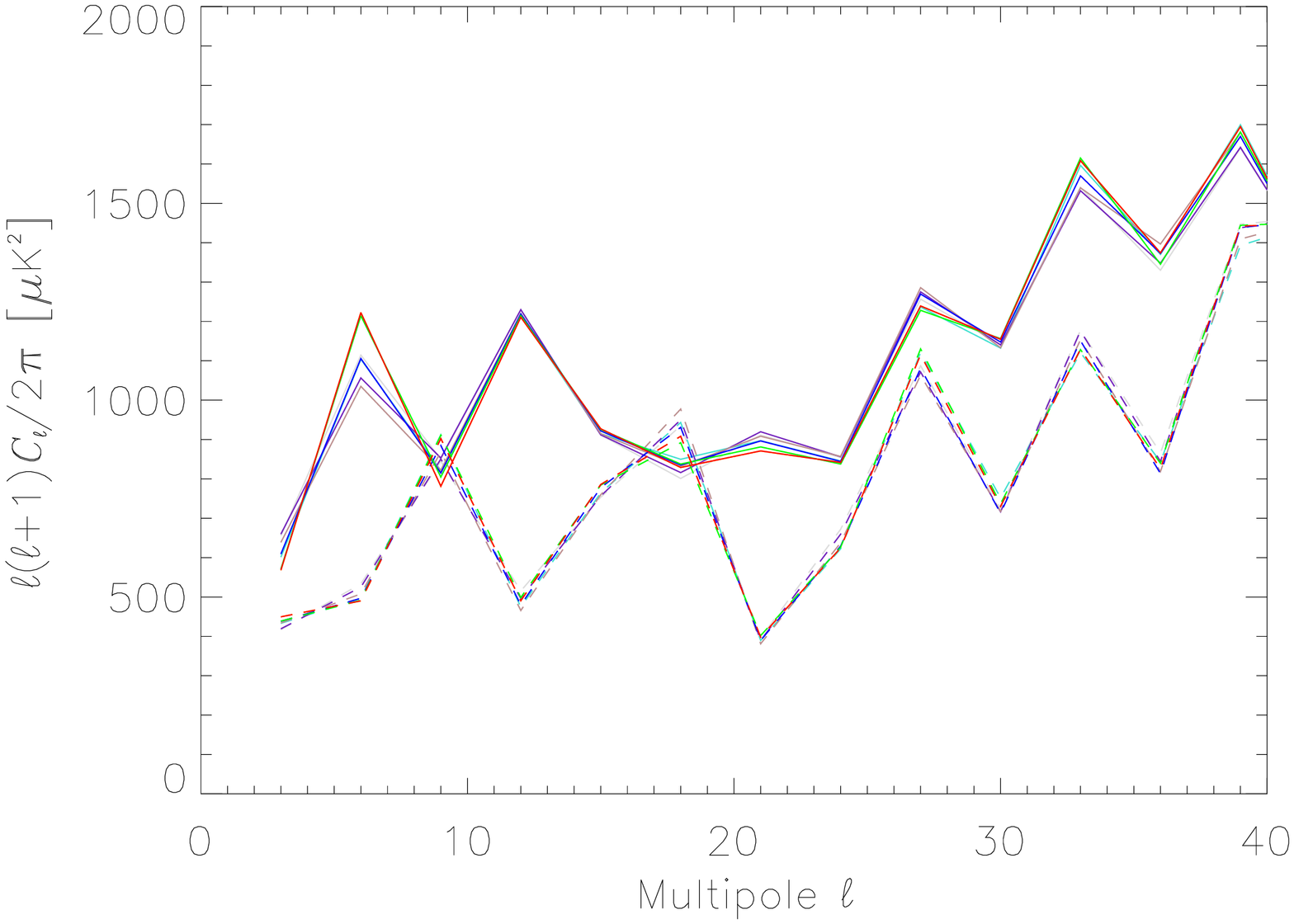}
\includegraphics[width=6.3cm,angle=90]{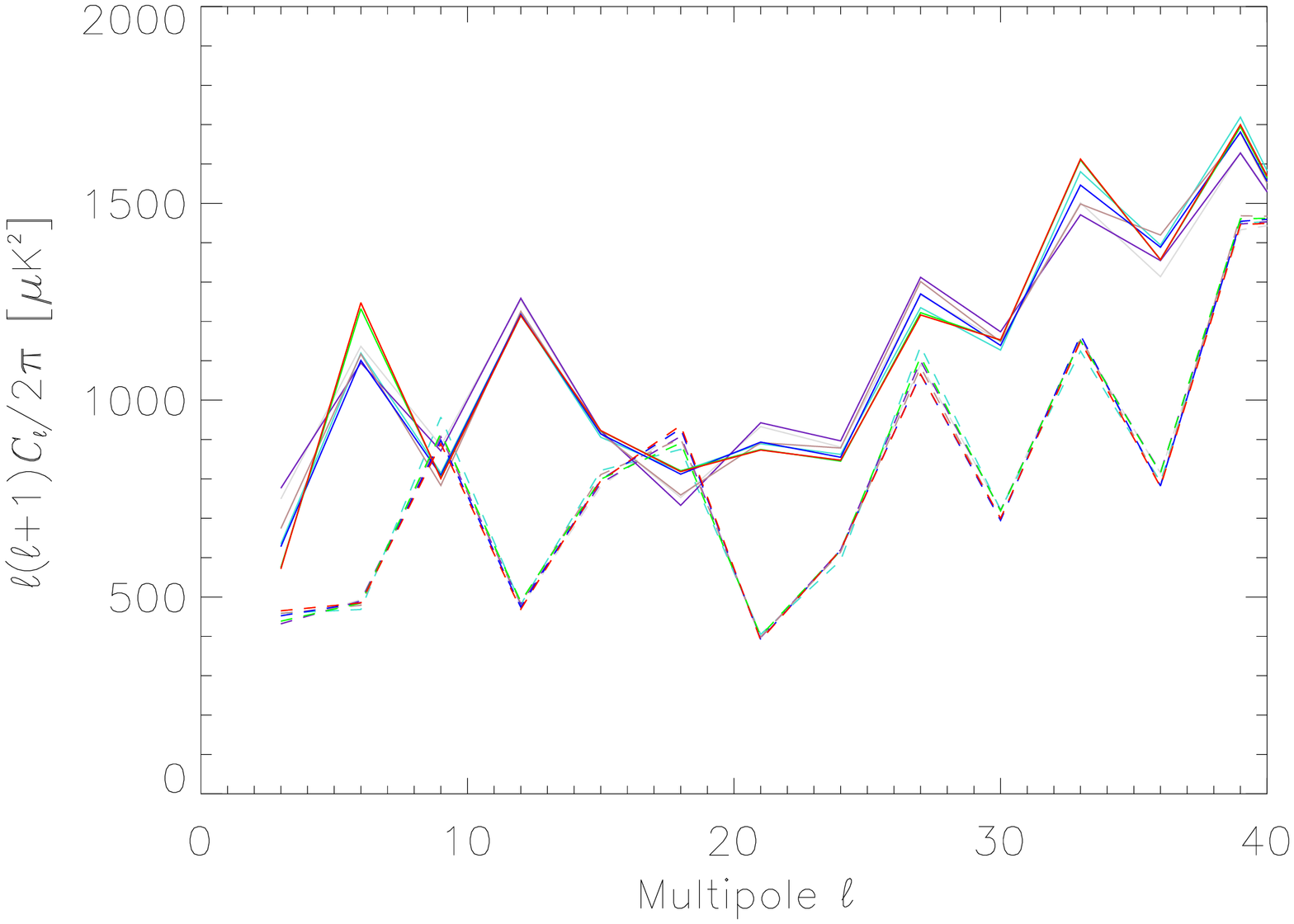}
\caption{The power spectra derived on the northern (dashed line) and southern
(solid line) hemisphere in the reference frame with north pole at
$(\theta,\phi)=(80^{\circ},57^{\circ})$. The left panel shows the spectra 
derived from the ICA CMB map out of Kp2. In the right panel the 
CMB maps are obtained applying \fstica~separately on the two hemispheres. Colors 
mark channel combinations as in previous figure.}
\label{ica_asym}
\end{centering}
\end{figure*}

As a first step, we computed the power spectrum on the 
ICA CMB maps, reconstructed out of Kp2, 
independently in the two hemispheres defined in the new reference frame. 
This is done for all the combinations of input frequency channels. 
The spectra are estimated in bins of 3 multipoles each from $\ell=2-40$
and adopting the same procedure described in the previous Section. The
results are shown in Fig.~\ref{ica_asym}~(left panel). We can see that for all
the channel combinations the southern spectrum has indeed more power than 
the northern one over almost the entire multipole range. We also performed
the same kind of analysis starting from maps obtained from the full sky
analysis and with Kp2 mask applied with identical findings.
In a pure \fstica~
analysis, one might be tempted to try an explanation in terms of a difference 
in the overall foreground spectral indices in the two hemispheres, since 
the \fstica~assumes an uniform frequency scaling across the whole sky; but the fact 
that other authors obtained the same result with totally independent procedures, 
and most importantly the test outlined below make this explanation unlikely. Indeed,  
we performed the component separation on the northern and
southern hemisphere separately and derived the ICA CMB power spectrum separately 
for each of them.
The spectra are reported in Fig.~\ref{ica_asym}~(right panel). Also in this case
the northern spectrum is systematically lower than the southern one.
This result strongly disfavors an explanation based on foregrounds for the asymmetry
found in the ICA CMB maps. In addition, we point out the remarkable 
agreement between our results and those of \citet{hansen_etal_2006} 
(see their Fig.~8), obtained with a completely independent technique. \\
On the \fstica~side, this confirms the reliability of the algorithm 
when exploited to reconstruct the finest structure in the CMB pattern out 
of a given dataset, as this work and the previous ones on BEAST 
\citep{donzelli_etal_2006} and \COBE~\citep{maino_etal_2003} demonstrate; 
on a purely scientific side, we confirm the existence of a marked 
asymmetry in the CMB anisotropy power between the considered northern and southern
hemispheres, which in this present case escapes explanations in terms of 
difference in the foreground properties on the corresponding two hemispheres. 

\section{Critical Discussion and Conclusion}
\label{conclusions}
 
In this work we applied the fast Independent Component Analysis (\fstica) 
component separation technique to the 3 years data of the Wilkinson Microwave 
Anisotropy Probe (\WMAP). The algorithm retrieves the different components 
superposed in a multi-frequency observation as linear combinations of the 
input data at different frequencies, which maximize the mutual statistical 
independence. We first evaluate the expected performance by means of Monte 
Carlo chains on simulated \WMAP~data varying the noise and CMB realization, 
and exploiting the existing foreground models. 
Among the recovered components, we identify one which is compatible with the CMB 
emission by checking the recovered frequency scaling. On simulations, the precision 
of the recovery on the frequency scaling is of the order of percent. We then 
apply the technique to the real data, again identifying the CMB component 
by means of the reconstructed frequency scaling. Different combinations of the 
\WMAP~channels give consistent results; in terms of the reconstruction of 
the CMB frequency scaling, the best \WMAP~configuration includes the three high 
frequency 
channels (Q, V and W, respectively at 41, 61, 94 GHz), plus the lowest one (K at 
23 GHz). This indicates that the algorithm 
benefits from having a good tracer of the low frequency foregrounds in the \WMAP~data in order to achieve a proper CMB cleaning. 
The recovered CMB power spectrum is in close agreement with the \WMAP~one on all 
accessible scales. The CMB fluctuation asymmetry in the northern and 
southern hemisphere claimed by several authors exploiting different data analysis 
techniques is confirmed in this work, with the same amplitude. 

The agreement of the present results on CMB with the \WMAP~ones and those from 
other authors is remarkable and strengthens the confidence we have on the CMB 
pattern reconstructed from the existing data as a whole. 
At the same time, from the point of view of the development of component separation 
techniques based on the Independent Component Analysis (ICA) this work represents the 
achievement of a most important milestone in view of the application to the 
Planck and other experiments; it means that the algorithm proved itself to be stable against the 
\WMAP~instrumental systematics, and realistic foregrounds. \\
On the other hand, we are aware of the limitations of the present analysis. First of 
all, \WMAP~is not an ideal experiment for performing a map based component 
separation, 
since the different channels have markedly different resolutions, and one has to do a 
pre-processing step decreasing the angular resolution of the data to a common one, 
which is about 1 degree in the present case. Second, the precision of the separation 
is evaluated by means of Monte Carlo chains on simulated data, which rely on ingredients, 
e.g. foregrounds and systematics, which may be far from reality, and also incomplete 
in the case of \WMAP, as the dust foreground is visible only in the W band at 
94 GHz, and does not have a multi-frequency coverage required by most component 
separation algorithms. 

Despite of all these oddities, the CMB pattern recovered 
by the \fstica~is consistent with the results in earlier literature, and 
confirms effects like the asymmetric distribution of the CMB anisotropy power 
across the sky.  This performance is likely to be due to the high level of 
statistical independence between background and foregrounds in the data, which is 
able to drive the solution close to the right one even for data affected by 
instrument systematics and realistic foregrounds. Thus we believe the present technique 
might be proving itself useful for future CMB probes, either accessing the CMB statistics 
beyond the power spectrum, and to control the foreground emission in polarization, 
which is going to be less known than in total intensity, and substantially higher 
compared with the CMB.

\section*{Acknowledgments}
Some of the results in this paper have been derived using the HEALPix 
\citep{gorski_etal_2005} package. This research was supported in part by the NASA LTSA grant NNG04GC90G.
DM acknowledge useful discussion with F.K. Hansen and thanks G.Hinshaw for
providing the full MASTER \WMAP~3yr power spectrum.
We acknowledge the use of the {\tt cmbfast} code and we thank the \WMAP~team
for making data available via the Legacy Archive for Microwave Data Analysis
(LAMBDA) at {\tt http://lambda.gsfc.nasa.gov}.

\label{lastpage}
\end{document}